\documentclass[12pt,preprint]{aastex}

\begin{document}

\title{On the Clustering of Sub-millimeter Galaxies}

\author{Christina C. Williams\altaffilmark{1}, 
Mauro Giavalisco\altaffilmark{1}, 
Cristiano Porciani\altaffilmark{2}, 
Min S. Yun\altaffilmark{1}, 
Alexandra Pope\altaffilmark{1},
Kimberly S. Scott\altaffilmark{3}, 
Jason E. Austermann\altaffilmark{4}, 
Itziar Aretxaga\altaffilmark{5}, 
Bunyo Hatsukade\altaffilmark{6},
Kyoung-Soo Lee\altaffilmark{7}, 
Grant W. Wilson\altaffilmark{1}, 
J. Ryan Cybulski\altaffilmark{1},
David H. Hughes\altaffilmark{5}, 
Ryo Kawabe\altaffilmark{6}, 
Kotaro Kohno\altaffilmark{8}, 
Thushara Perera\altaffilmark{9}, 
F. Peter Schloerb\altaffilmark{1}}

\altaffiltext{1}{Astronomy Department, University of Massachusetts, 710 North Pleasant Street, Amherst, MA 01003, USA, ccwillia@astro.umass.edu}
\altaffiltext{2}{Argelander-Institut f\"ur Astronomie der Universit\"at Bonn, Auf dem H\"ugel 71, D-53121 Bonn, Germany}
\altaffiltext{3}{Department of Physics and Astronomy, University of Pennsylvania, 209 South 33rd Street, Philadelphia, PA 19104, USA}
\altaffiltext{4}{Center for Astrophysics and Space Astronomy, University of Colorado, Boulder, CO 80309, USA}
\altaffiltext{5}{Instituto Nacional de Astrofisica, «Optica y Electr«onica (INAOE), Aptdo. Postal 51 y 216, 72000 Puebla, Pue., Mexico}
\altaffiltext{6}{Nobeyama Radio Observatory, Minamimaki, Minamisaku, Nagano 384-1805, Japan}
\altaffiltext{7}{Yale Center for Astronomy and Astrophysics, Department of Physics, Yale University, New Haven, CT 06520, USA}
\altaffiltext{8}{Institute of Astronomy, the University of Tokyo, 2-21-1 Osawa, Mitaka, Tokyo 181-0015, Japan}
\altaffiltext{9}{Department of Physics, Illinois Wesleyan University, Bloomington, IL 61701, USA}

\begin{abstract}
We measure the angular two-point correlation function of sub-millimeter galaxies (SMGs) from 1.1-millimeter imaging of the COSMOS field with the AzTEC camera and ASTE 10-meter telescope. These data yields one of the largest contiguous samples of SMGs to date, covering an area of
0.72 degrees$^{2}$ down to a 1.26 mJy/beam (1-$\sigma$) limit, including  189 (328) sources with S/N$\ge3.5$ (3). We can only set upper limits to the correlation length $r_{0}$, modeling the correlation function as
a power-law with pre-assigned slope.  Assuming existing
redshift distributions, we derive $68.3\%$ confidence level upper
limits of $r_{0}\lesssim6$-$8~h^{-1}$Mpc at 3.7 mJy, and $r_{0}\lesssim11$-$12 ~h^{-1}$Mpc at 4.2 mJy. Although consistent with most previous estimates, these upper limits imply that the real $r_{0}$ is likely smaller.  This casts doubts on the robustness of claims that SMGs are characterized by significantly stronger spatial clustering, (and thus larger mass), than differently selected galaxies at high-redshift.  Using Monte Carlo simulations we show that even strongly clustered distributions of galaxies can appear unclustered when sampled with limited sensitivity and coarse angular resolution common to current sub-millimeter surveys. The simulations, however, also show that unclustered distributions can appear strongly clustered under these circumstances.  From the simulations, we predict that at our survey depth, a mapped area of two degrees$^{2}$ is needed to reconstruct the correlation function, assuming smaller beam sizes of future surveys (e.g. the Large Millimeter Telescope's 6$''$ beam size). At present, robust measures of the clustering strength of bright SMGs appear to be below the reach of most observations.

\end{abstract}


\section{Introduction}

High-redshift galaxies which are relatively bright at millimeter and
sub-millimeter wavelengths, and thus detectable by current ground--based
instrumentation, have, over the last decade, come to the forefront of studies
of galaxy evolution. Commonly referred to as sub-millimeter galaxies (SMGs),
because the first significant deep surveys have been made at $\lambda=450$ and
850 $\mu$m, these sources are thought to be largely obscured by dust, with
star-formation rates of up to 1000 $M_{\sun} {\rm year}^{-1}$ needed to power
their high rest-frame infrared luminosity of $L_{IR} \sim 10^{12}-10^{13}
L_{\sun}$ \citep{Smail1997, Hughes1998, Barger1998}. It has long been
speculated that with such high luminosity and star formation rates, SMGs
should be very massive, strongly clustered, and trace large-scale structure at
high redshift \citep{Blain2004, Amblard2011}. If it is conclusively found that
SMGs do cluster strongly in space, and therefore trace massive dark matter
halos at high redshift, this will provide additional evidence that these
sources are evolutionarily linked to massive elliptical galaxies often found
in the center of galaxy clusters in the local universe \citep{Lilly1999,
  Eales1999}. Hence, robust determination of the clustering strength of the
SMGs, at least of those that are commonly detected with current
instrumentation, namely with flux brighter than a few mJy, would have strong
implications in theories of galaxy evolution \citep{VanKampen2005,
  Negrello2007}, as it is not well understood what observable properties of
galaxies are characteristic of biased tracers of the background dark matter
distribution.

Until recently, a secure measurement of SMG clustering has been elusive
\citep{Webb2003, Blain2004, Scott2006, Weiss2009}, in large part
because of the slow mapping speeds of sub-millimeter instruments, whose maps
have been very small in area. Recently, data at 250-500$\mu$m from the 
Herschel Space Telescope, has produced improvements in terms 
of the area of sub-millimeter maps, allowing clustering measurements 
to be made with improved statistics \citep{Maddox2010, Cooray2010}. 
The Herschel surveys, however, are biased to low redshift and low luminosity 
galaxies as a result of the bluer wavelengths that they cover and of
the negative k-correction; they also still suffer from 
source confusion despite their large area. Clustering measurements at 
longer wavelengths, on the other hand, still remains uncertain. Sub-millimeter 
surveys are still
limited by large beam size and shallow survey depths which have prevented
robust positions and large sample sizes. As sub-millimeter maps become larger
with higher resolution, and the number of securely detected sources becomes
statistically significant, studies of clustering of these SMGs will surely
provide interesting implications for galaxy evolution.

Here we present measures of the angular clustering of SMGs from one of the
first millimeter maps containing a statistically significant number of
SMGs. This map covers a 1 square degree section of the Cosmic Evolution Survey
\citep[COSMOS;][]{Scoville2007} field using the AzTEC bolometer array, mounted
on the Atacama Sub-millimeter Telescope Experiment (ASTE). The data provide
the largest contiguous map, and the largest galaxy sample, at 1.1 millimeters
to date (Aretxaga et al., in prep). While spectroscopic redshift information
on SMGs remains sparse, we have used various redshift distributions to
estimate de-projected spatial clustering for these galaxies. We assume a
cosmology with $\Omega_{\Lambda} = 0.7$, $\Omega_{m} = 0.3$, and $H_{o} = 100
h$ km ${\rm s}^{-1}$ ${\rm Mpc}^{-1}$.

\section{Observations}

The COSMOS field was mapped with AzTEC on ASTE, a full description and results
will be presented in a separate publication (Aretxaga et al., in prep.). We
imaged a subset of the COSMOS blank field centered at $(RA, Dec) = (150.125,
2.23)$ with a total area of 1.41 degrees$^{2}$, totaling 112.6 hours of
observing time.
With AzTEC on ASTE, the beam size is $28''$ (full-width half max). 
For this analysis we have considered only the region of the map where the
coverage was $50\%$ of the maximum value or higher. This results in a
contiguous map of 0.72 degrees$^{2}$. We achieve an average noise level of
1.26 mJy/beam, which is very uniform throughout the area considered, ranging
from 1.23 to 1.27 mJy/beam.

\begin{figure}[!ht]
\centering
\includegraphics[scale=.7]{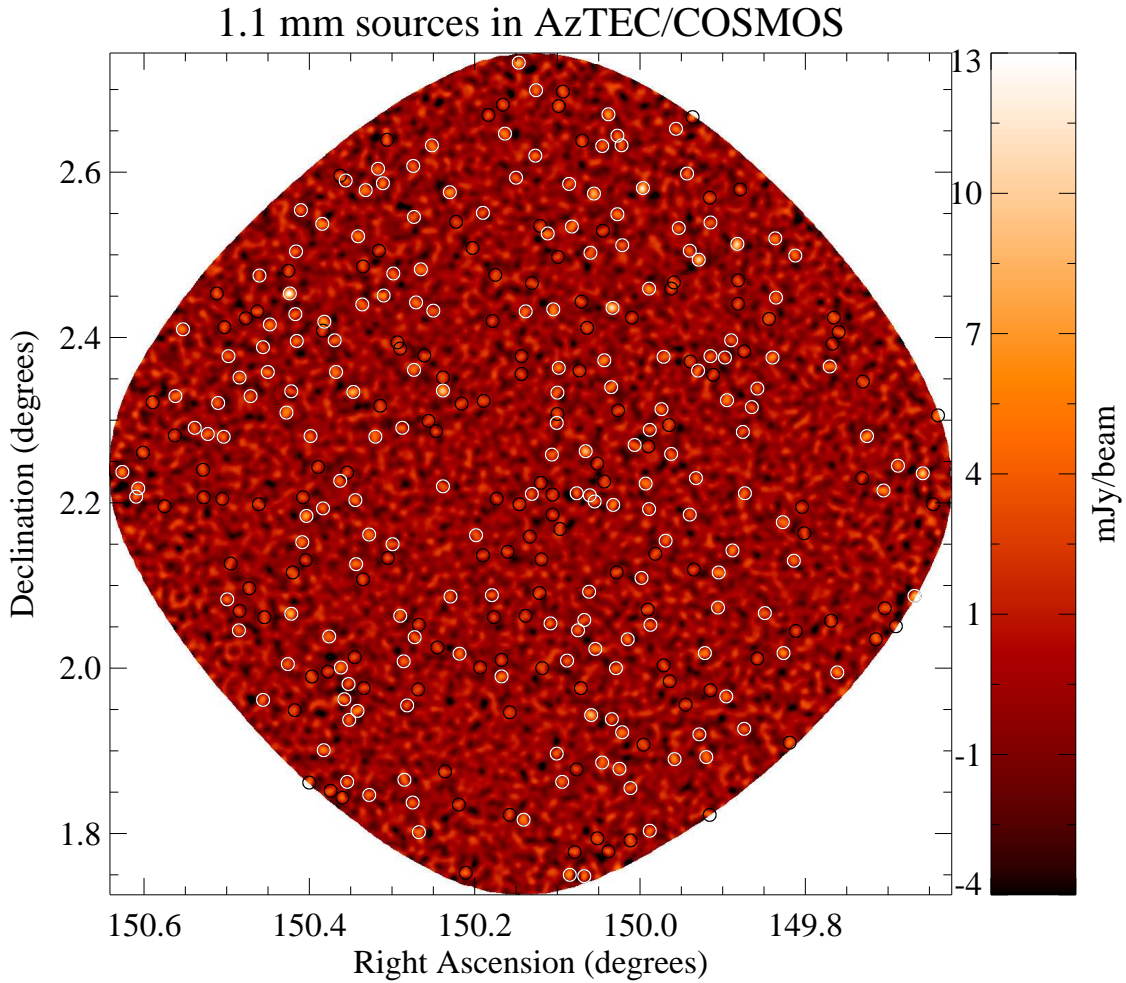}
\caption[]{The $50\%$ coverage region in the AzTEC/COSMOS map. 3.0 to
  3.5-$\sigma$ sources are circled in black, $>$3.5-$\sigma$ sources are
  circled in white. Circle size corresponds to one and a half times the
  beam size. \label{map}}
\end{figure}

Our millimeter sources are selected by searching for peaks above a given
signal-to-noise ratio (S/N) with a window corresponding to the beam size
\citep[e.g.][]{Scott2008}. We find 328 sources with a S/N above 3.0, and 189
sources with a S/N above 3.5, hereafter the 3.0-$\sigma$ and 3.5-$\sigma$
catalogs, respectively. The map and source positions are shown in Figure
\ref{map}.

\section{Clustering Analysis}
\subsection{Angular Clustering}

\begin{figure}[!ht]
\centering
\epsscale{.90}
\plotone{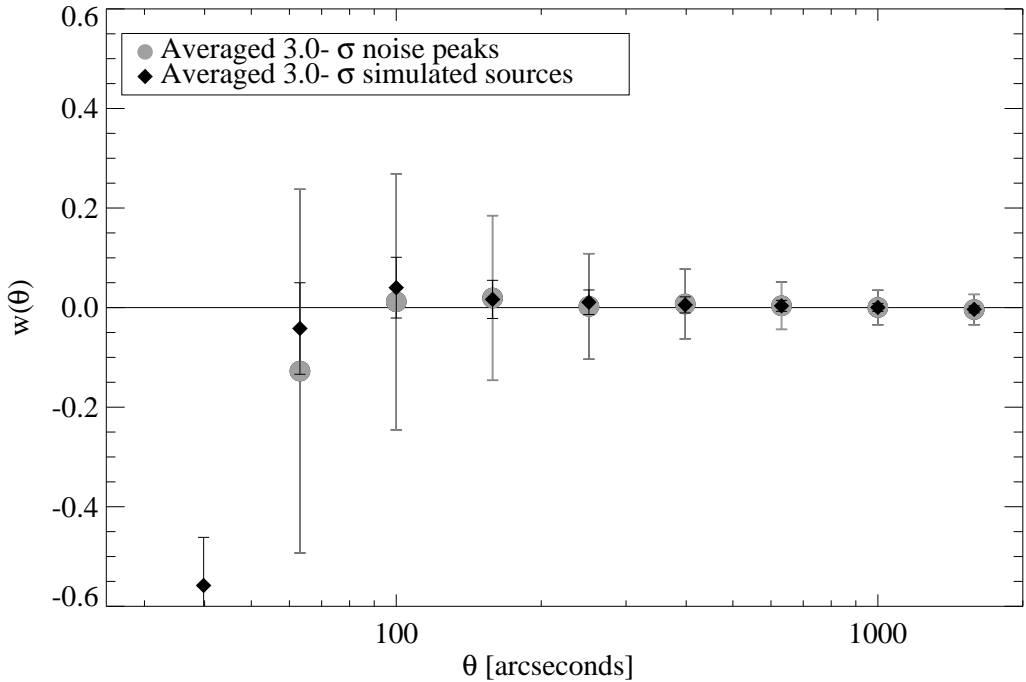} % noise_acf.eps
\caption[]{Averaged ACF for the peaks detected at 3.0-$\sigma$ from the 100 noise realizations (gray circles) and for the simulated sources detected at 3.0-$\sigma$ (black diamonds). Errors indicate standard deviation. Gray horizontal line corresponds to zero clustering. Slight anti-correlation around the 60'' bin is a result of the beamsize; in general more random pairs with angular separations in this bin will be found since distances between detected sources are not smaller than twice the beamsize.\label{noisefig}}
\end{figure}

The angular two-point correlation function, $w(\theta)$, measures the excess
probability, above that expected for a random distribution, of finding two
galaxies with an angular separation $\theta$, within a solid angle $\delta
\Omega$. It is defined in terms of the probability $\delta P =
N^{2}[1+w({\theta})]\delta \Omega$, where $N$ is the surface density of galaxies
\citep{Peebles1980}. We measure angular clustering of SMGs in the COSMOS field
using the Landy-Szalay estimator of the angular correlation function
\citep[ACF;][]{LandySzalay1993}. This can be measured as 
$$w(\theta) = \frac{DD(\theta)-2DR(\theta)+RR(\theta)}{RR(\theta)},$$ where
$DD(\theta)$ is the number of observed galaxy pairs as a function of angular
separation, $\theta$, $DR(\theta)$ are the number of cross-pairs between the
observed galaxies and a randomly distributed sample, and $RR(\theta)$ is the
number of randomly distributed pairs. The random distributions are generated
by inserting randomly positioned sets of artificial sources into realizations
of the noise distribution in our COSMOS map. The injected sources have a flux
distribution based on our best estimate of SMG number counts from blank field
observations \citep{Austermann2010}, but we tuned the parameters such that the
number of significant sources retrieved by our source finding algorithm are on
average within $2\%$ of the number of detections in the real map. We generate
100 of these simulations, and we use the random sources selected above the
corresponding S/N threshold from each as random distributions. We have
verified that (on average) the noise peaks in our COSMOS map and sources in
the simulations are unclustered at the angular separations we consider (see Figure \ref{noisefig}).

We expect our uncertainty to be dominated by small number statistics, but it
is possible that the map properties, such as non-uniformity and beam size, contribute to our error in measuring the
ACF. So rather than assume Poisson errors\footnote[1]{We found that Poisson errors are sometimes incorrectly used in
  the literature, as $\delta
  w(\theta)=\sqrt{\frac{1+w(\theta)}{DD(\theta)}}$. The correct expression is
  $\delta w(\theta)=\frac{1+w(\theta)}{\sqrt{DD(\theta)}}$
  \citep{LandySzalay1993}. } 
given by $\delta w(\theta)=\frac{1+w(\theta)}{\sqrt{DD(\theta)}}$ \citep{LandySzalay1993},
which do not take these effects into account, we also
quantify the uncertainty using the simulations. To do this we calculate the
ACF of each of the 100 simulated random catalogs, whose intrinsic ACF we know (on average we found that
simulated sources are unclustered). Thus, the standard deviation of the ACF of the individual simulated catalogs, $\sigma_{sim}(\theta)$, should include any error in $w(\theta)$ related to the properties of the map. This uncertainty is given by $\delta w(\theta)=(1+w(\theta))\sigma_{sim}(\theta)$.
We find the uncertainty obtained this way to be smaller than the Poissonian
error for both catalogs, so we have conservatively assumed the latter.

\begin{figure}[!ht]
\includegraphics[scale=.3,angle=270]{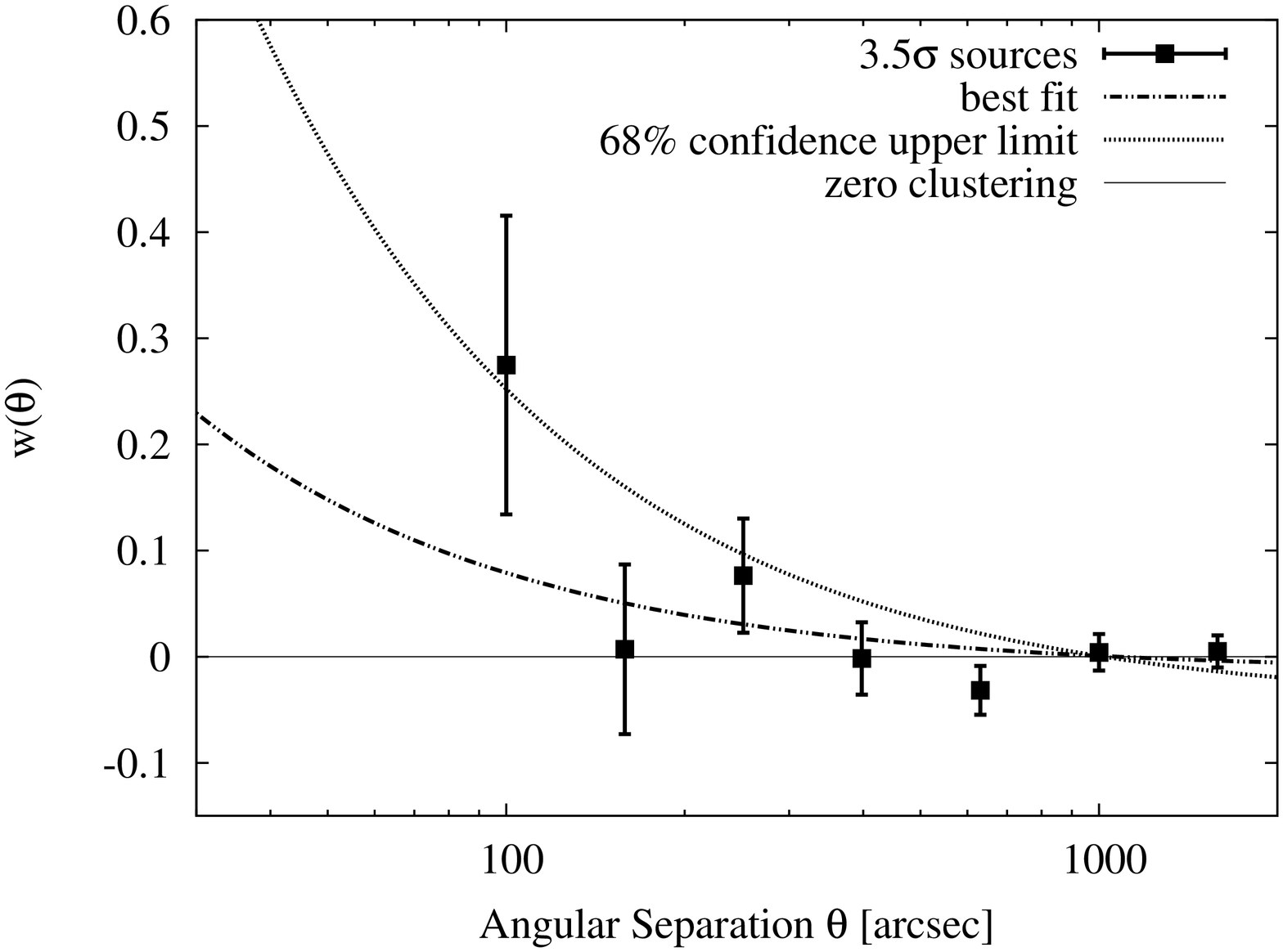} % smg_acf35v9_poi.eps
\includegraphics[scale=.3,angle=270]{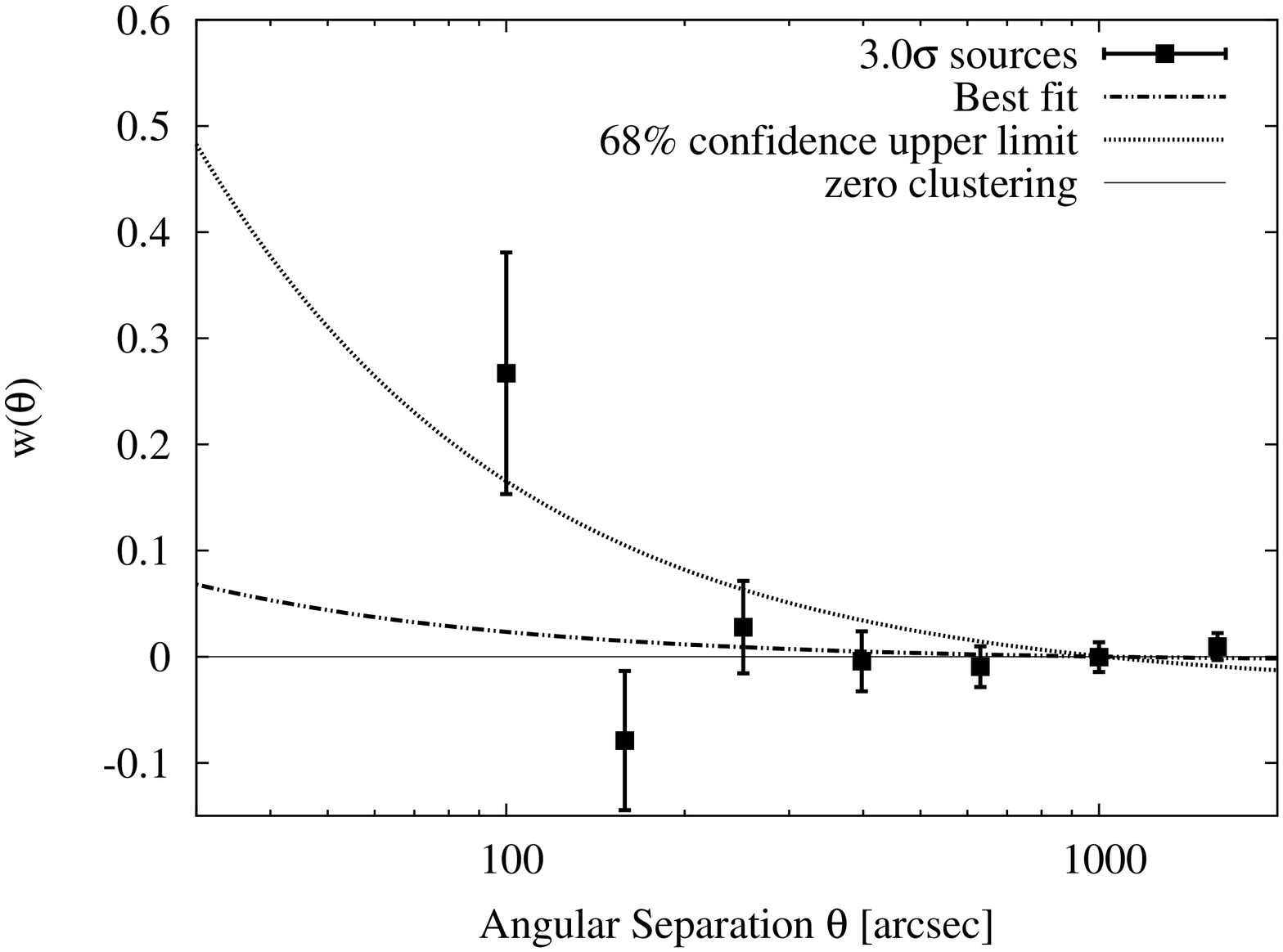} % smg_acf30v9_poi.eps
\caption[]{ACF for the two catalogs (points). The lower dot-dashed lines are
  the best fit power-law to the data, upper dotted lines are the $68.3\%$
  confidence level upper limits. Both power laws shown assume $\beta = 0.8$,
  and have their corresponding IC subtracted to match what is fit to the
  data. \label{fig3}}
\end{figure}

\begin{figure}[!ht]
\centering
\epsscale{.90}
\plotone{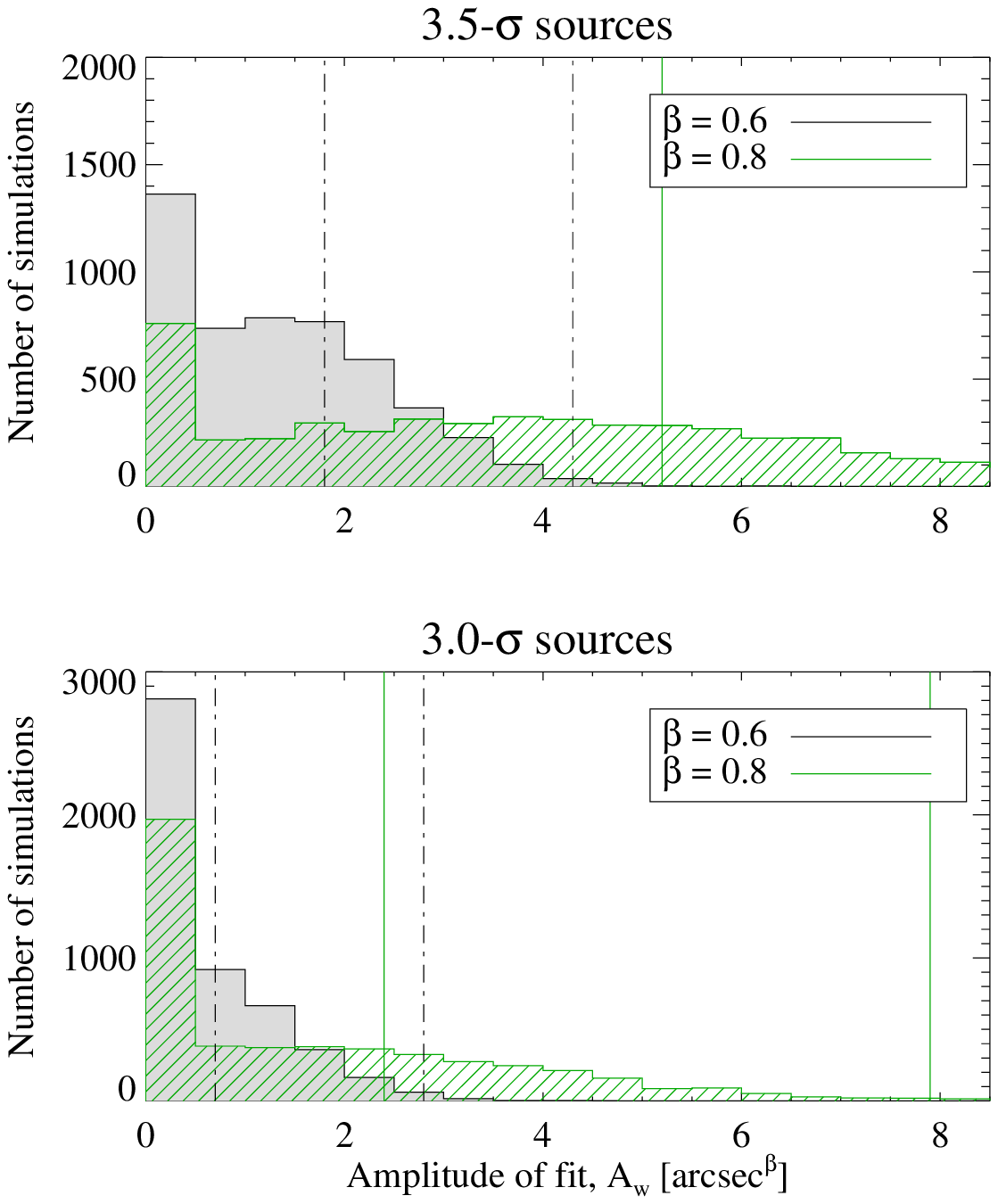} % awv6_b6dec16.eps
\caption[]{Distributions in amplitudes for power law fits with slopes fixed to
  $\beta = 0.8$ and $0.6$ for 3.0- and 3.5-$\sigma$ catalogs, generated by
  Monte Carlo simulations of the observed ACF.  The vertical lines correspond
  to the $68.3\%$ (smaller $A_{w}$) and $99.5\%$ (larger $A_{w}$) limits in
  each distribution, and represent the largest clustering strengths allowable
  by our datasets. See Table 1 for corresponding values. \label{fig4}}
\end{figure}

The difference between the 3.0- and 3.5-$\sigma$ catalogs from the observed
map is a trade off between a larger catalog of galaxies, and a lower false
detection rate of sources in our observed map.
By setting a high S/N threshold for detection we can achieve negligible 
false detections, but our number of sources would be small. We choose to select
sources with a lower S/N threshold, while acknowledging that some fraction of
them are not real SMGs. For the 3.5-$\sigma$ catalog, we expect that about
$9\%$ of the sources are false detections, and for the 3.0-$\sigma$ catalog,
$24\%$ are false detections (Aretxaga et al, in prep.).  Including some
fraction of randomly positioned non-galaxies will only serve to dilute our
estimate of the clustering.  As noise peaks are inherently unclustered, we can
correct for this effect as $w_{obs}(\theta) = (1-f)^{2}w_{true}(\theta)$ where
$f$ is the fraction of false detections included in the catalog. Our ACF
estimate, corrected for dilution by random false detections, is presented in
Figure \ref{fig3}. We do not include in the analysis ACF measurements at
angular separations smaller than twice the beam size.

We assume that at the angular separations we are considering, the ACF behaves
as a power law of the form $w(\theta) = A_{w}\theta^{-\beta}-IC$, where we
refer to $A_{w}$ as the clustering amplitude. IC refers to the integral
constraint correction, which we calculate using the algorithm of
\cite{Roche1999}. The result of a power law fit where the clustering amplitude
and slope $\beta$ are left as free parameters is poorly constrained, and the
best slope is unphysically steep due to the fact that the measured ACF is high
at the lowest angular scale. So given our large uncertainties, we do not
attempt to constrain both clustering amplitude and slope. Instead, we assume
two different representative values for $\beta$. The first, $\beta = 0.8$, is
observed for massive elliptical galaxies in large low redshift surveys and is
the value typically assumed for massive galaxies and SMGs at high redshift
\citep{Zehavi2002, Blain2004}. The second value, $\beta = 0.6$, is the
shallower slope typically observed for normal ultra-violet selected
starbursting galaxies at high redshifts such as Lyman-break galaxies (LBGs),
and BX/BM galaxies \citep{GiavaliscoDickinson2001, PorcianiGiavalisco2002,
  KSLee2006, Adelberger2005}, as well as starforming galaxies at low redshift
\citep{Zehavi2002}. Unless otherwise stated, in the text we will quote results
derived using $\beta = 0.8$.

In Figure \ref{fig3} we also show the best fitting power laws (assuming $\beta
= 0.8$) for each catalog found from a least squares minimization.  Due to the
large uncertainties, the best fit amplitudes are poorly constrained and
the 1$\sigma$ upper limits to these best fit values are large. These upper limits are also
shown in Figure \ref{fig3}.  The case of zero clustering lies within the
$1\sigma$ error (defined as $\Delta \chi^{2} < 1$), but as negative values imply anti-correlation and are considered unphysical, we set zero clustering to be the lower limit. These best fits, 1-$\sigma$ upper error, and lower error (as described above) are $A_{w} = 3.7^{+8.2}_{-3.7}$ arcsecond$^{0.8}$ for the 3.5-$\sigma$
catalog and $A_{w}=1.1^{+6.7}_{-1.1}$ arcsecond$^{0.8}$ for 3.0-$\sigma$, and
are summarized in Table \ref{table1} along with results assuming $\beta =
0.6$.

While this $\chi^{2}$ analysis provides a best fit with confidence intervals, it does not limit a priori the possible range of values that $A_{w}$ can assume. We want to explore the effect on the power law fit if we only consider positive values for $A_{w}$, as negative values are unphysical. To do this we perform Monte Carlo simulations where we generate 5000 Gaussian deviated realizations of the observed ACF. The Gaussian deviates are generated using
the Poisson error on each value of $w(\theta)$. We fit each deviated
realization with the same power law form outlined above to produce a
distribution of best fitting clustering amplitudes, $A_{w}$. The resulting
distributions in $A_{w}$, given assumed values of $\beta$, are shown for each
catalog in Figure \ref{fig4}. For both catalogs, the most likely values to be
measured for $A_{w}$ is zero, corresponding to the case where SMGs are
unclustered. It must be emphasized that this does not mean that SMGs such as
these are spatially unclustered, only that the strength of their clustering is
below what is robustly detectable from our survey. The peak at zero is merely an effect of the likelihood of the $\chi^{2}$ distribution extending into the negative values of $A_{w}$, since $A_{w}$ is poorly constrained. 

Using these distributions shown in Figure \ref{fig4}, we set upper limits to the power law amplitudes
which are allowable given our measured ACF for SMGs, so our results can be
compared with previous measurements of SMG clustering. These distributions are one-sided (because the peak lies at zero, the lowest value we allow for $A_{w}$), and so the distributions can only provide an upper limit. This is in contrast to the $\chi^{2}$ distribution which is two-sided and so provides an upper and lower limit (where conventionally the $68.3\%$ confidence limits are given by $\Delta \chi^{2}< 1$). The two upper limits are different from each other in that the $68.3\%$ confidence level upper limit from the $\chi^{2}$ minimization corresponds to a $15.85\%$ probability of finding a larger $A_{w}$, whereas the one-sided $68.3\%$ confidence level upper limit from the Monte Carlo simulation corresponds to a $31.7\%$ probability of finding a larger $A_{w}$. We find that using these distributions from the Monte Carlo simulations we can
reject power law models with amplitudes larger than $A_{w} = 2.4$
arcsecond$^{0.8}$ at the $68.3\%$ confidence level, and $A_{w} = 7.9$
arcsecond$^{0.8}$ at the $99.5\%$ confidence level for the 3.0-$\sigma$
catalog, and $A_{w} = 5.2$ and $A_{w} = 11.5$ arcsecond$^{0.8}$ at the
$68.3\%$ and $99.5\%$ confidence levels, respectively, for the 3.5-$\sigma$
catalog. These results are shown as solid and dashed vertical lines in Figure
\ref{fig4} and are summarized in Table \ref{table1}.

%\begin{table*}[!ht]
\begin{table}[!ht]
\begin{center}
\caption{SMG Clustering Results.\label{table1}}
\begin{tabular}{cccccccccccccccc}
\tableline\tableline
Catalog & N & $S$\tablenotemark{a} & Fit Type\tablenotemark{b} & $\beta$ & $A_{w}$\tablenotemark{c}  & & IC\tablenotemark{d} &
$r_{o}$\tablenotemark{e}  & & 
$r_{o}$\tablenotemark{f}  & \\
 & & [mJy] & & & best & upper & & best & upper  & best & upper \\ 
 & &  & & & &  limit& & & limit &  &   limit\\ 

\tableline
% as of dec 16, 2010: correct sims:
3.5-$\sigma$ & 189 & 4.2 & $\chi_{\nu}^{2}$ & 0.8 & 3.7 & 11.9 & .015 & 10.0  & 19.2 & 9.6 & 18.4 \\
 & & & & 0.6 & 1.2 & 4.4 &  .018 & 9.5 & 21.4 & 9.0 & 20.4 \\
&  &  & MC $68.3\%$ & 0.8   & & 5.2  &   & & 12.1 & & 11.6 \\
 &  &  &  &  0.6 & & 1.8 &  & & 12.2 & & 11.7 \\
 &   &  & MC $99.5\%$  & 0.8  & & 11.5  &   & &  18.8 & & 18.1 \\
 &   &  &  &  0.6 & & 4.3 &  &  & 21.1 & & 20.1 \\
\tableline
3.0-$\sigma$ & 328 & 3.7 & $\chi_{\nu}^{2}$ & 0.8 & 1.1 & 7.8  & .004 & 5.1 & 15.2 & 4.7 &  14.6 \\
 & & & & 0.6 & 0.3 & 2.9 & .004 & 4.0 & 16.5 & 3.8 & 15.7 \\
 & & & MC $68.3\%$ &  0.8 & & 2.4 & & & 7.9 & & 7.6 \\
 & & &  & 0.6 & & 0.7 &  & & 6.8 & & 6.5 \\
& & & MC $99.5\%$ & 0.8 & &7.9 & & & 15.3 & & 14.7 \\ 
& & & & 0.6  &  &2.8 & & & 16.1 & & 15.4 \\

\tableline
\end{tabular}

\tablenotetext{a}{Flux limit at 1.1 mm of the catalog} \tablenotetext{b}{For
  the reduced $\chi^{2}$ fit, best fitting results are listed with the
  corresponding upper limit where $\Delta \chi_{\nu}^{2} > 1$.  In the case of
  the Monte Carlo results (indicated by MC), values are percentage of
  confidence level upper limit in the acceptable value of amplitude $A_{w}$,
  given the SMG catalog.}  \tablenotetext{c}{Best fits and upper limits to
  $A_{w}$ (in arcseconds$^{\beta}$). Lower limits in all cases are zero as
  explained in the text.}  \tablenotetext{d}{IC values correspond to the best fit power-law.} \tablenotetext{e}{Correlation length in units
  $\ h^{-1}$Mpc, given our assumption of redshift distribution of
  \cite{Chapman2005}} \tablenotetext{f}{Using redshift distribution of
  \cite{Chapin2009}}
\end{center}
%\end{table*}
\end{table}

\subsection{Spatial Clustering}

To derive the spatial correlation length we have de-projected the angular
correlation function using the Limber transformation \citep{Peebles1980} and
assuming a redshift distribution for SMGs. Robust measures of this
distribution are limited, in large part because coarse
angular resolution of sub-millimeter and millimeter maps results in large
positional uncertainties, making counterpart identification for spectroscopic
followup difficult. The crude knowledge of the redshift distribution for these
galaxies provide an additional source of systematic error in the
derivation of the spatial clustering. Here we discuss results obtained by
assuming two redshift distributions which are believed to be representative of
SMGs detected in the same range of far--IR wavelengths as the ones considered here. The most
widely used is the distribution of \cite{Chapman2005}, compiled from a set of
75 spectroscopic redshifts for 850$\mu$m-selected SMGs with optical
counterparts identified using deep interferometric radio continuum
imaging. This redshift distribution is known to be biased towards low
redshifts due to the requirement of a radio detected counterpart, so we use
the version of this distribution which has been corrected for the radio
bias. The corrected distribution is well described by a Gaussian peaking at $z
= 2.3$ and a spread of 1.2, ranging from $1 < z < 3.5$. Using this
distribution, we find that the $68.3\%$ confidence level upper limit of
consistent correlation lengths for SMGs are $\lesssim 6$-$8 \ h^{-1}$Mpc and
$\lesssim 11$-$12 \ h^{-1}$Mpc for the 3.0-$\sigma$ and 3.5-$\sigma$ catalogs,
respectively.  Results assuming the redshift distribution of \cite{Chapin2009}
produce similar values, which are summarized in Table \ref{table1}.  The
\cite{Chapin2009} redshift distribution is generated from a combination of
spectroscopic and photometric redshifts for 1.1-mm detected galaxies, so may
be more applicable to this study. The distribution differs from that of
\cite{Chapman2005} in that it peaks around $z=2.7$ and has a high-redshift
tail to $z\gtrsim4$. We emphasize that these results are upper limits, and
therefore the intrinsic clustering of this set of galaxies are likely to be lower.

\begin{table*}
\begin{center}
\caption{SMG Clustering and Flux Limits.\label{table3}}
\begin{tabular}{ccccccccc}

\tableline\tableline
$N_{sources}$ & $r_{o} \pm \delta r_{o}$ & $\lambda$ & $S_{\nu}$ & $S_{850 \mu
  m}$ & beam size  & Reference  &  \\
 & [$\ h^{-1}$Mpc] & [$\mu$m] & [mJy] & [mJy] & [arcseconds] &\\

\tableline
27 &$12.8 \pm 7.0$ & 850  & 3.0  & 3.0  & 14.5 & Webb et al. 2003 \\
47 & $6.9 \pm 2.1$  & 850  & 5.0  & 5.0  & 14.5 & Blain et al. 2004 \\
51 & 31 \tablenotemark{a} & 850  & 5.0 & 5.0 & 14.5 & Scott et al. 2006 \\
126 &$13 \pm 6$ & 870   & 4.6  & 4.8\tablenotemark{b} & 19.2 & Weiss et al. 2009   \\
1633 & $7-11$ & 350  & 36 & 7.3\tablenotemark{b} & 17 & Maddox et al. 2010\\
189 &  $<$11-12\tablenotemark{c} & 1100  & 4.2  & 6.7\tablenotemark{b} & 28 & This study\\
328 &  $<$6-8\tablenotemark{c} & 1100  & 3.7 & 5.9\tablenotemark{b} & 28 & This study\\

\tableline
\end{tabular}

\tablenotetext{a}{Only angular clustering was published by \cite{Scott2006}, we transform their power-law result and errors for their sources above S/N of 3.5, using the redshift distribution of \cite{Chapman2005}}
\tablenotetext{b}{Flux density translated assuming $S_{\nu} \propto \nu^{1.8}$} 
\tablenotetext{c}{We have quoted our $68.3\%$  confidence level upper limits for comparison.}
\end{center}
\end{table*}

\section{Discussion}
\subsection{Comparison to other SMG clustering measurements}

Before comparing our clustering measures with other works it is important to
keep in mind the limitations inherent in the selection of samples based on an
observable property, such as flux, as opposed to a physical property, such as
luminosity or mass. The term ``Sub--Millimeter Galaxies'' is often used to
indicate a category of galaxies (population is the term often used in this
context) thought to have well specified and somewhat homogeneous
properties. For example, SMGs are commonly interpreted as massive systems
characterized by prodigious star--formation rates powered by major merger
events. While these properties most likely apply to {\it some} SMGs of
relatively large far--IR luminosity, it obviously is unreasonable to think
that they are generic to {\it any} galaxy that is detectable at some
wavelengths around 1 mm. Firstly, we should remind that galaxies detected at
some wavelength with some telescope/instrument combination do not, generally
speaking, span the same range of the far--IR luminosity function or redshift
as galaxies from another instrumental configuration observed at another
wavelength in the sub-millimeter/millimeter spectral region. Lumping all such
samples as ``SMG'' believing that they share very similar properties is
misleading. In other words, the definition of ``SMG'' as galaxies that are
detected at wavelengths crudely in the range 500 $\mu$m to 1 mm at
the sensitivity of current survey facilities, does not result into the
selection of common physical properties. It is true that, since current
ground--based facilities working at the popular 850 $\mu$m wavelength have
limited dynamic range in sensitivity, the resulting samples of galaxies at
similar redshifts also have similar far--IR luminosity and thus, presumably,
physical properties. But this is just an ``observational accident'' that does
not apply to other sub--millimeter surveys. In general, galaxies detected at
350 $\mu$m with {\it Herschel}/SPIRE or at 1.1-mm with ASTE/AzTEC, even if at
the same redshift as those observable with JCMT/SCUBA, will cover different
portion of the far--IR luminosity function, and will generally have different
physical properties, such as mass, clustering strength, star--formation rate,
etc. (we are not addressing here the different sensitivity and redshift
distribution function of the corresponding samples).

With this caveat in mind, we can try to compare our results with others.  We
find that at 1.1mm and down to 1.26 mJy the angular clustering of SMGs in the
COSMOS field is poorly constrained and with our sample size we can only set
upper limits to the correlation length. Our $68.3\%$ confidence level upper
limits to the correlation length from the Monte Carlo simulation are
$\lesssim6$-$8 \ h^{-1}$Mpc or $\lesssim11$-$12 \ h^{-1}$Mpc, depending on flux
limit. Generally, our SMG clustering limits are higher for the higher flux
limit (4.2 mJy). The recent prediction from the theoretical model of SMG
clustering by \cite{Almeida2010} for $S_{850\mu m} > 5$mJy is $r_{o} \sim 5
\ h^{-1}$Mpc, which is consistent within the error of our measurements. The
$S_{\rm 1.1mm} > 3.7$mJy flux limit of the 3.0-$\sigma$ sources roughly translates
to $S_{850\mu m} > 5.9$mJy, assuming the spectral index suggested by
\cite{Chapin2009} of $S_{\nu} \propto \nu^{1.8}$ between 1.1-mm and 850-$\mu$m
sources. Our 1.1-mm sources are likely similar to this simulated galaxy
distribution.

\begin{figure}[!ht]
\centering
\plotone{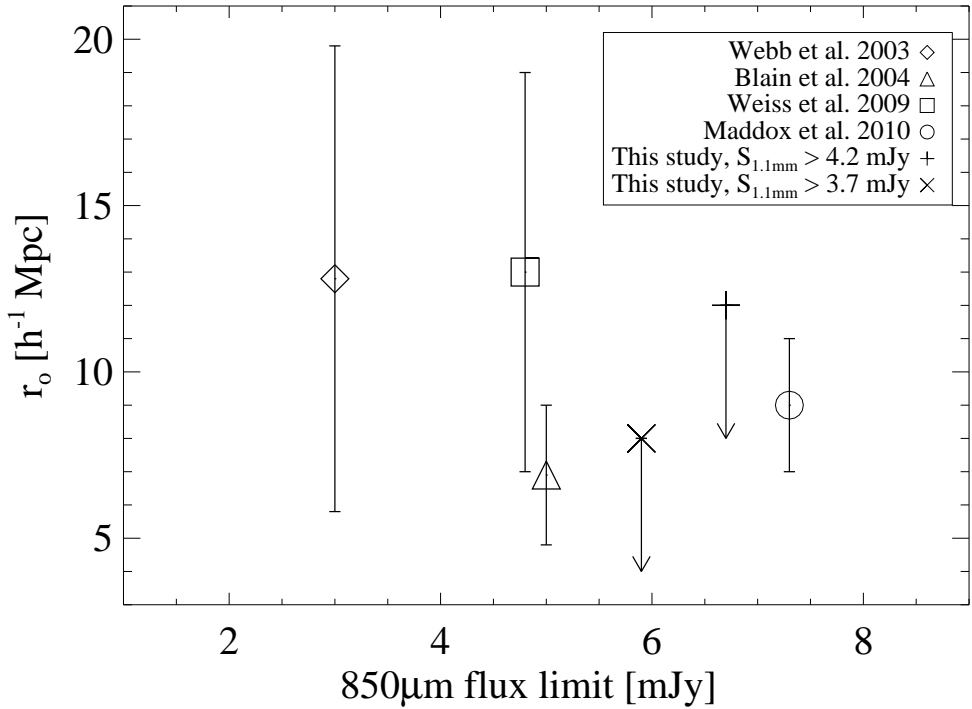} % fluxlim1.eps
\caption[]{Correlation lengths and flux limits (translated to 850$\mu$m using the spectral index of Chapin et al. (2009)) from this and previous studies. 
 \label{fluxlim}}
\end{figure}

In terms of the results of previous studies of SMG clustering, direct
comparisons are difficult to make because of the differing wavelengths and
flux limits of each survey.  Poorly quantified redshift distributions in all
cases further complicate the issue when trying to make comparisons. As we
found in this paper, the brighter SMGs show evidence for stronger clustering
than the fainter SMGs consistent with what is expected from galaxy evolution
models (e.g.~Almeida et al.~2010). A similar result is also found by
\cite{Brodwin2008} based on the clustering of {\it Spitzer}-selected
ultra-luminous infrared galaxies. This demonstrates that caution should be
taken when comparing the clustering of different samples of SMGs selected at
different wavelengths down to different depths. If we ignore these effects, a
direct comparison makes our upper limits inconsistent with the results of
\citep{Scott2006}, but consistent with the majority of other previous studies
due to large uncertainties which tend to be larger than 20-50 percent in
$r_{o}$ \citep{Webb2003, Blain2004, Weiss2009, Maddox2010}. These comparisons
are summarized in Table \ref{table3} and Figure \ref{fluxlim}. Our flux limit
is generally higher than other studies, when translated to a common
wavelength, so it may be reasonable to assume the AzTEC sources should be more
strongly clustered.

In a recent paper, Amblard et al.~(2011) measured the clustering at
250--500$\,\mu$m from the brightness fluctuations in the power spectrum of
{\it Herschel}/SPIRE maps after masking out the bright, detected sources.
Assuming we are probing the Raleigh-Jeans tail of the spectral energy
distribution, the average flux density ratio is $S_{\rm 1.1mm}/S_{350\mu m}\sim8$
(with a spectral index of 1.8, Chapin et al. 2009). Given the confusion limit
of SPIRE, these fluctuations are probing the clustering of sources down to
350$\,\mu$m fluxes of a few mJy (Amblard et al.~2011), which translates to
only 0.4 mJy and 0.25 mJy at 850$\,\mu$m and 1.1mm, respectively. This limit
probes galaxies down to L$_{\rm{IR}}\sim3\times10^{11}\,L_{\odot}$ at
$z\sim2$, much fainter than the typical limits of sub-millimeter surveys. It
is not clear which part of the far-IR luminosity function contributes most to
the clustering signal measured by the fluctuations.

While the Amblard et al.~(2011) result provides an interesting constraint on
the clustering of fainter sub-millimeter-emitting galaxies, these sources are
much more numerous than typical SMGs \citep[e.g.][]{Smail2002}, and are not
expected to evolve into the most massive elliptical galaxies in the local
Universe. The halo masses derived in Amblard et al.~(2011) are more comparable
to those of the less extreme Lyman Break galaxies than the bright, detected
SMGs. With larger telescopes such as the LMT and CCAT, we will be able to
individually detect galaxies down to $S_{\rm 1.1mm}<0.1\,$mJy and measure the
clustering as a function of luminosity, a strong test of various galaxy
evolution models.

The strength of SMG clustering is an additional test of evolutionary models because it can discriminate between the various formation mechanisms for SMGs. Discriminating between merging or cold-mode accretion as the dominant mechanism by which SMGs form at high-redshift is of particular interest, and recent simulations of each mechanism predict different correlation lengths. The model of \cite{Dave2010}, where SMGs are formed by accretion of large amounts of cold gas, predicts a large correlation length ($r_{o} \sim 10\ h^{-1}$Mpc) because cold gas accretion should be most influential in the most massive dark matter halos. Merger driven scenarios on the other hand predict a more modest range in correlation lengths, between $r_{o} = 5$-$6 \ h^{-1}$Mpc \citep{Almeida2010}. We are not yet at the point where we can see distinguishing evidence between the models, but this will also be an important goal of larger sub-millimeter observatories.

Additionally, due to the large uncertainty in our measurement, our results are
also consistent with measurements of weaker clustering from other types of
high-redshift starforming galaxies such as LBGs and other restframe-UV
selected galaxies, BzKs, and unresolved sources contributing to the cosmic
infrared background \citep{KSLee2006, Adelberger2005, GiavaliscoDickinson2001,
  Hayashi2007, Viero2009}. Their minimum halo masses of $\sim10^{11}$-$10^{12}
M_{\sun}$ and correlation lengths of about $r_{0} \sim 4$-$5 \ h^{-1}$Mpc
\citep{KSLee2006, PorcianiGiavalisco2002, Adelberger2005} are consistent with
the masses and correlation lengths for both bright and faint SMGs
\citep{Almeida2010, Amblard2011}. If the underlying sub-millimeter galaxy
population we detected in this study is weakly clustered, as may be implied by
\cite{Almeida2010} and \cite{Amblard2011}, it supports our conclusion from
section 2 that the clustering is too weak to be detected with our survey.

\subsection{Map limitations on measuring clustering}

From a practical point of view, an important question to answer is: what
characteristics of area and depth should surveys of SMGs have in order to
yield robust measures of clustering. For example, how much area and down to
which flux limit, should a survey with AzTEC reach in order to test the
hypothesis that SMGs at the bright end of the far-IR luminosity function are
the progenitors of massive elliptical galaxies, and should therefore be
strongly clustered? In addressing this question, one needs to take into
account the key contributors to the error budget of the measures, such as 1)
the uncertainty in the redshift distribution, since a wide one that covers a
large redshift interval washes out the clustering signal due to projection
effects; 2) the sparse sampling of the underlying SMG population, which
determines the shot noise in the ACF measures; 3) the large beam size of
current observations, which prevents one from measuring the ACF at small
angular scales where the signal is strongest.

We have done Monte Carlo simulations to investigate the extent to which these map
properties are affecting our ability to measure SMG clustering. Specifically,
we measure the ACF from realizations of galaxy distributions for which we have
defined the intrinsic clustering, and impose AzTEC-like map properties. The
realizations are made by generating a log-normal density distribution with an
intrinsic ACF, and Poisson sampling the density field according to the methods
outlined in \cite{PorcianiGiavalisco2002}. The resulting realizations are
$0.72 deg^{2}$ in area and contain on the order of $10^{4}$ mock galaxies. To
match the expected percentage of false detections, we merge the clustered mock
set with a set of random positions, so that they make up $9\%$, like the
3.5-$\sigma$ catalog. We then randomly sample points from the realization to
match the observed number density of 3.5-$\sigma$ sources in the AzTEC map,
where the sampled objects are never closer than one beam size separation, and
see how their ACFs compare with the intrinsic ACF of the realization.  We test
intrinsic ACFs which are strongly and weakly clustered according to power-laws
of $A_{w}=2.9$ and $A_{w}=0.5$, respectively, where $\beta =0.8$. These
correspond to values of $r_{o} \sim 9 \ h^{-1}$Mpc and $r_{o} \sim 4\ h^{-1}$Mpc
for our assumed redshift distribution function. We again disregard ACF
measurements for angular separations smaller than twice the beam size. The
purpose of this test is to simulate the ACF we should expect to observe from a
map similar to the AzTEC-COSMOS map, if SMGs are intrinsically strongly or
weakly clustered galaxies.

\begin{figure}[!ht]
\centering
\plotone{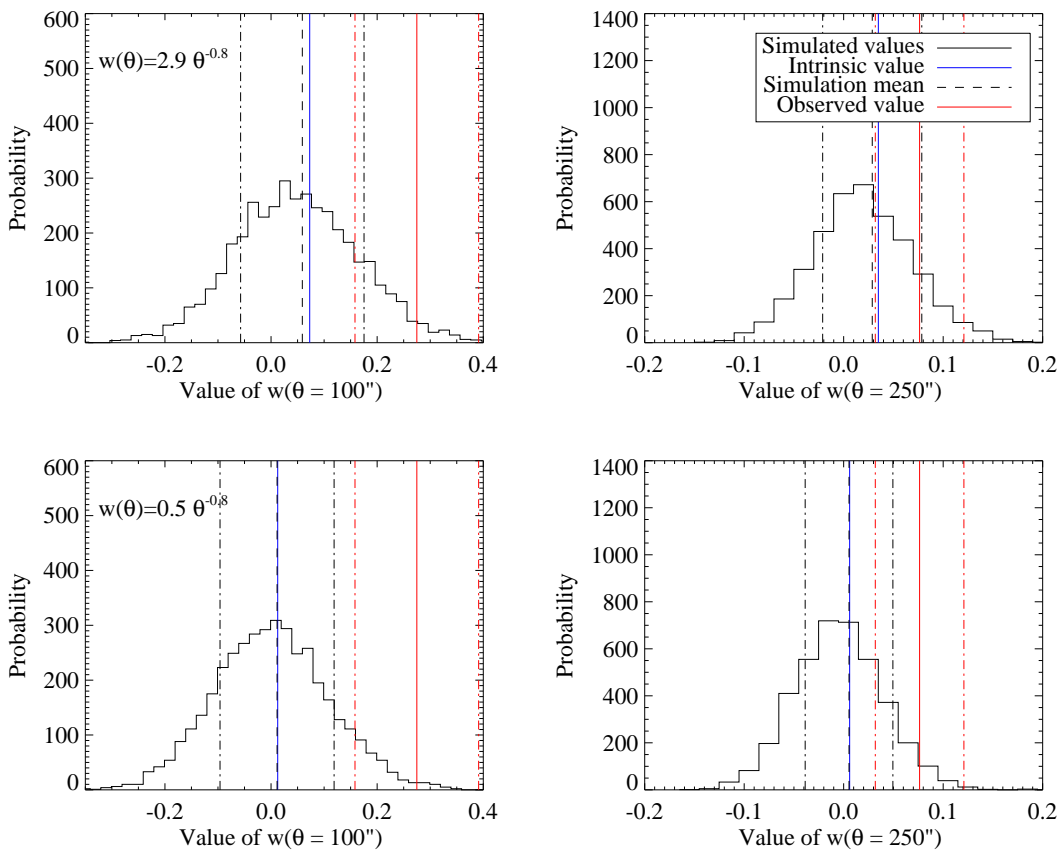} % mockcatsimang.eps
\caption[]{Probability distributions for the ACF at 100 and $250''$
  separations from the mock catalogs (black histogram). Dashed line is the
  mean from the mock catalogs, which when corrected for the IC agrees well
  with the intrinsic value (blue line). Histograms are roughly Gaussian, with
  standard deviation indicated by dot-dashed lines. Solid red is the observed
  ACF from the 3.5-$\sigma$ catalog at those angular separations, with Poisson
  errors given by red dot-dashed lines. Each row are results from an intrinsic
  power-law form shown in the left panel. \label{simfigang}}
\end{figure}

\begin{figure}[!ht]
\centering
\epsscale{0.86}
\plotone{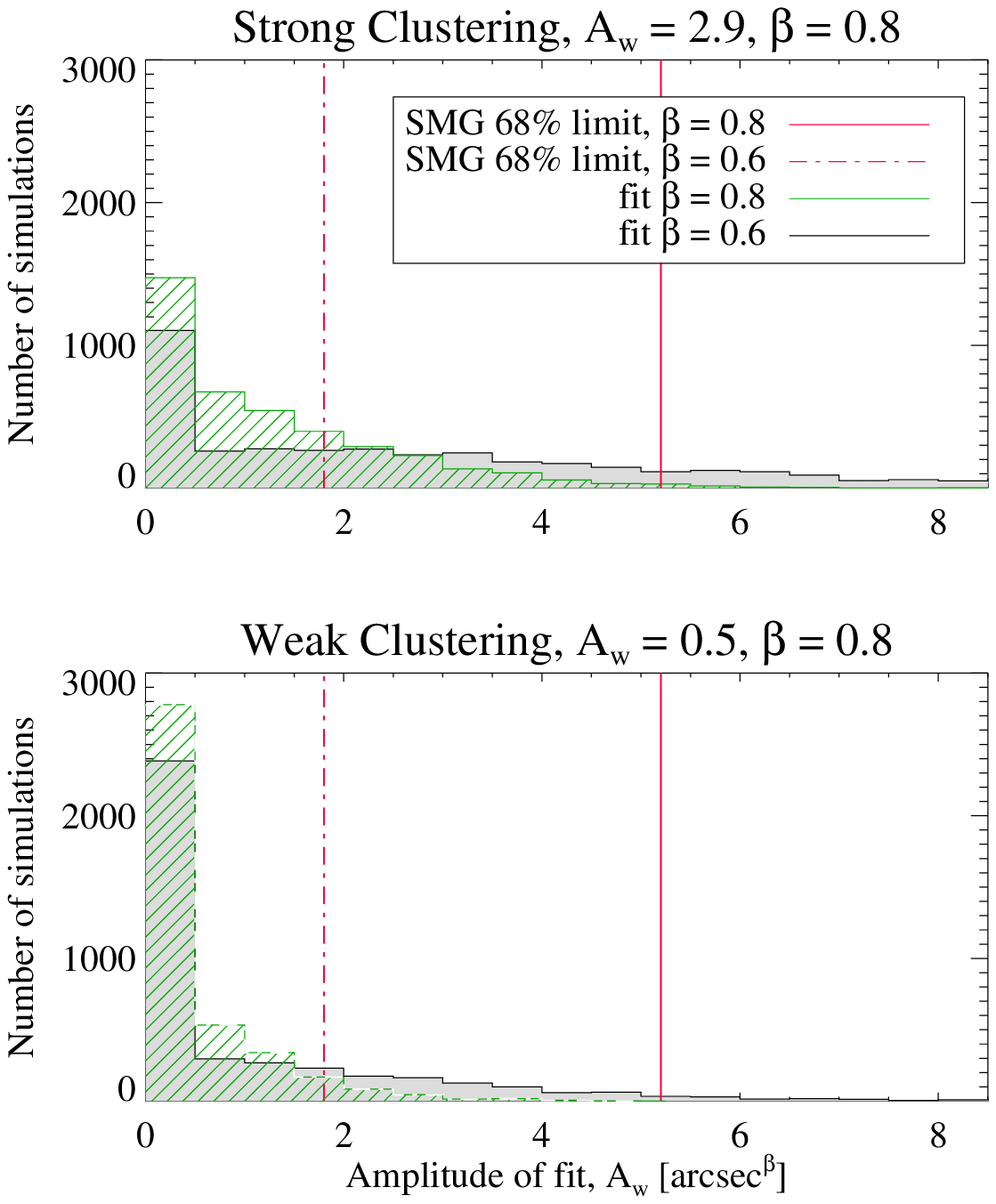} % mockcatsimnew.eps
\caption[]{The distributions of clustering amplitudes from fitting power-laws
  to ACFs from clustered simulated maps. Top panel is results from strong
  intrinsic clustering ($A_{w}=2.9$), and bottom is from weak clustering
  ($A_{w}=0.5$), where $\beta = 0.8$. Green hatched histograms are $A_{w}$
  distributions when fitting with an assumed $\beta = 0.8$, and in gray are
  assuming $\beta=0.6$. Red lines are the $68.3\%$ upper limits in $A_{w}$
  from the fit to the ACF of the 3.5-$\sigma$ AzTEC sources, assuming
  $\beta=0.8$ (solid red), and $\beta = 0.6$ (dot-dashed red).
 \label{simfig}}
\end{figure}

In Figure \ref{simfigang}, we show the distributions in the value of $w(\theta
= 100'')$ and $w(\theta = 250'')$ from the simulations for each power-law form
we tested, after applying the correction for false detection rate. The
distributions of the ACF values are very broad as should be expected because
of the low number of mock galaxies used in each sample, as well as the
fraction of random positions. If the measured values are corrected for the
integral constraint then the peak of the distributions at each angular
separation match well with the intrinsic value. When the distributions are
compared to our observation, there is a very small chance of spuriously
finding the observed value at $\theta = 100''$ for the 3.5-$\sigma$ sources,
about $3\%$ for the strong case and $0.75\%$ for the weak case. At larger
separations the chance increases to $31\%$ and $15\%$, respectively. However,
the observed and simulated values for both angular separations for the strong
clustering case are consistent within their errors.

We have fit a power-law to the mock ACFs, assuming a fixed $\beta$ of 0.6 or
0.8. The clustering amplitudes we recovered are shown in Figure \ref{simfig},
along with the upper limits we derived from the real AzTEC map. These
histograms are essentially probability distributions for clustering amplitudes
that will be measured from $189$ sources in the AzTEC map area if the
intrinsic population is strongly or weakly clustered. In none of the cases
explored here is it likely that the intrinsic power-law form will be
recovered. The assumed $\beta$ influences the shape of the distribution, but
the probability always peaks at zero. There is $22\%$ chance that fits
assuming $\beta = 0.8$ will indicate zero clustering ($24\%$ for $\beta =
0.6$) even if the intrinsic correlation length is $r_{o}= 9 \ h^{-1}$Mpc. If the
intrinsic value is $r_{o} = 4 \ h^{-1}$Mpc, the percentages increase to $53\%$
and $55\%$ assuming $\beta = 0.8$ and 0.6, respectively. Even though the
distributions in Figure \ref{simfigang} nicely correspond with the intrinsic
value, the power-law distributions in Figure \ref{simfig} peak at zero because
the large fluctuations in each realization from the small sample size can
cause negative values in the ACF.  It is not possible to recover the intrinsic
clustering properties, and it is not possible to differentiate between strong
and weak clustering. The implications this has for millimeter and
sub-millimeter surveys at this resolution and sensitivity, or any survey where
such a sparse sampling of the underlying population is detected, is that the
true clustering properties cannot be recovered.

\subsection{Predictions for future surveys}

\begin{figure}[!ht]
\centering
\plotone{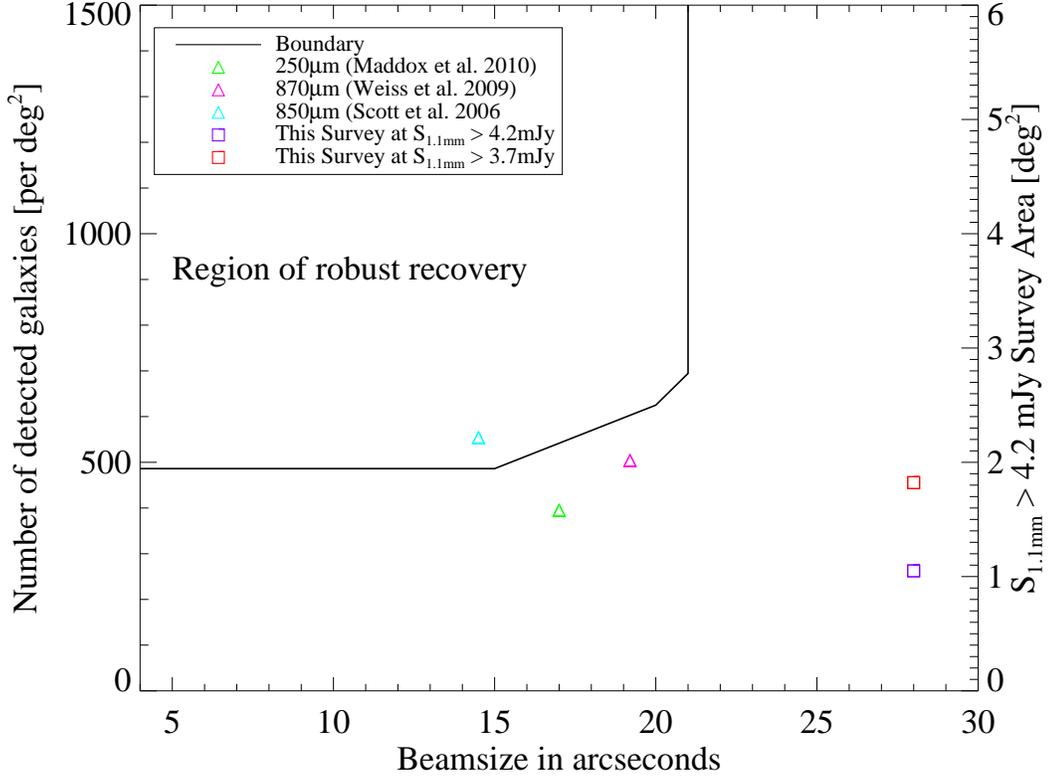} % NvsB.eps
\caption[]{Galaxy sample number needed to recover an intrinsic clustering with
  power-law form $w(\theta)=2.9\theta^{-0.8}$, as a function of beam size. Area
  up and left from the contour line indicates region where values may be
  recovered. Right-side axis indicates survey area required at this
  sensitivity to robustly measure the clustering of the $S_{\rm 1.1mm} > 4.2$ mJy
  sources investigated here. Some positions of previous surveys have been
  provided where possible, based on the source number density of their survey.
% However due to their sample having a mix of sources whose positions have
% been measured with and without identified counterparts, we have excluded the
% \cite{Blain2004} study from this comparison.
\label{NvsB}}
\end{figure}

Our sensitivity to the clustering signal in this study is determined by two
things. First, the number of sources, which depends on the area mapped and the
depth, must be large enough to overcome the high shot noise stemming from the
sparse sampling of the underlying galaxy distribution by the SMG
selection. Second, the ability to measure small-scale separations between
galaxies, which depends on the beam size. Not surprisingly, we generally found
that the probability to recover the intrinsic ACF increases slightly with
decreasing beam size, however the intrinsic value for a sample size such as
ours was still not the most likely to be observed down to a beam size of
$5''$. Increasing the number of detected galaxies, which can be achieved by
increasing survey sensitivity or survey area, provides the largest
improvement. Fortunately, millimeter and sub-millimeter facilities are
advancing and future studies of SMG clustering will benefit from increased
sample sizes and improved angular resolution. Thus the real question becomes,
what resolution and survey area will be necessary to get an accurate measure
of SMG clustering?  Using the strongly clustered simulation discussed in the
previous paragraph for an ASTE-COSMOS sized map, we have estimated the
limiting galaxy sample size (as a function of beam size) for which it is
possible to recover the intrinsic clustering. A measurement of the clustering
is considered to have recovered the intrinsic value if, after sampling the
parent realization 2000 times and fitting the ACFs, the value of the intrinsic
clustering amplitude lies within the standard deviation of the clustering
amplitude distribution. Additionally, the distribution must satisfy the
requirement that the most likely value of clustering amplitude in the
distribution also lies within the standard deviation. This second condition
rejects the types of distributions shown in Figures \ref{fig4} and
\ref{simfig}. The resulting 'region of robust recovery' is shown in Figure
\ref{NvsB}. We have added the approximate positions of previous surveys which
have made clustering measurements, assuming the estimated number density of
detected sources in each is constant. These placements indicate that
previously measured ACFs, even the Herschel surveys with large area [16 square
  degrees], are still compromised by large beams and low sensitivity. The
previous study by Scott et al (2006), measured from a combination of multiple
SCUBA fields, falls barely within the robust region. However, one caveat of
this simulation is that it assumes a contiguous map region. In this case, the
signal to noise of the ACF measurement, which depends on the number of
independent galaxy pairs, DD, is related to the total number of detected
galaxies N by $DD = 0.5N(N-1)$. For discontiguous fields, galaxy pairs between
fields cannot contribute and so the number of galaxy pairs is lower, given the
same number of detected galaxies. For discontiguous identical fields, the number of pairs goes down by a factor of F where F is the number of fields. Thus, measurements made from galaxies in multiple fields will inevitably have lower signal to noise than a measurement using the same number of galaxies from a contiguous area.

The region in Figure \ref{NvsB} illustrates the difficulties in measuring the
angular clustering of bright SMGs such as the $S_{\rm 1.1mm} > 3.7$ mJy samples
explored here, but additionally provides some guidance for future surveys
which will aim to robustly measure the clustering of SMGs. Upcoming surveys
with the Large Millimeter Telescope for example, with its beam size of 6$''$,
will be able to make robust measurements for these galaxies with a mapped area
of about two square degrees. These future results will no doubt provide
exciting discoveries about the parent population of sub-millimeter sources.

\section{Summary}

1. We have measured the angular clustering of SMGs detected at 1.1mm from the
largest contiguous map at that wavelength to date. We have studied sources
detected at 3.5 (3)-$\sigma$ with flux limits $S_{\rm 1.1mm} > 4.2 (3.7)$ mJy. The
power-law fits are poorly constrained due to large uncertainties in the ACF.

2. We have set upper limits to the spatial correlation lengths for these
galaxies. For flux limits $S_{\rm 1.1mm} > 4.2$ mJy, we find $r_{0}\lesssim
11$-$12 \ h^{-1}$Mpc, and for $S_{\rm 1.1mm} > 3.7$ mJy we find $r_{0}\lesssim 6$-$8
\ h^{-1}$Mpc.

3. We have shown that for simulated clustered samples, our map properties,
specifically survey area, depth, beam size, prevent us from accurately
measuring strong clustering (e.g. with $r_{0}\sim9 \ h^{-1}$Mpc).

4. We have used these simulations to predict the conditions under which future
surveys may robustly detect clustering. Specifically, to measure clustering
galaxies detected with $S_{\rm 1.1mm} > 4.2$ mJy and mapped to a depth of 1.26
mJy/beam, we will be able to robustly measure clustering with an area of
$\sim2$ square degrees, with the LMT's beam size of 6$''$.

\acknowledgments

The authors wish to thank Sara Salimbeni, and Paolo Cassata for useful discussions, and Dan Popowich for technical support. We also thank Asantha Cooray for valueable comments. Support for this work was provided in part by the National Science Foundation grants AST-0838222 and AST-0907952.

\bibliographystyle{aa}
%\bibliography{ccw_v9}

\end{document}